\documentclass[final,5p,times]{elsarticle}

\usepackage{amsmath}          
\usepackage{amssymb}          
\usepackage{mathrsfs}         
\usepackage{mathtools}        

\usepackage{etoc}             
\usepackage{soul}             
\usepackage{parcolumns}

\usepackage{caption}          
\usepackage{makecell,multirow,multicol,booktabs} 
  
\usepackage{enumitem}         
\setlist[enumerate]{topsep=2pt,parsep=0pt,itemsep=0pt}
\setlist[itemize]{topsep=2pt,parsep=0pt,itemsep=0pt}

\usepackage{tikz}
  \usetikzlibrary{calc,positioning,shapes.geometric}
\usepackage{pgfplots}
  \pgfplotsset{compat=newest}
\definecolor{jr@lightblue}{RGB}{209,229,240}
\definecolor{jr@lightgreen}{RGB}{233,255,233}
\definecolor{jr@orange}{RGB}{255,127,0}
\definecolor{jr@pink}{RGB}{247,129,191}
\definecolor{jr@purple}{RGB}{152,78,163}

\usepackage{listings}
\definecolor{codegreen}{rgb}{0,0.6,0}
\definecolor{codegray}{rgb}{0.5,0.5,0.5}
\definecolor{codepurple}{rgb}{0.58,0,0.82}
\definecolor{backcolor}{rgb}{0.95,0.95,0.92}
\lstdefinestyle{mystyle}{
  backgroundcolor=\color{backcolor},
  commentstyle=\color{codegreen},
  keywordstyle=\color{magenta},
  numberstyle=\tiny\color{codegray},
  stringstyle=\color{codepurple},
  basicstyle=\ttfamily\footnotesize,
  breakatwhitespace=false,
  breaklines=true,
  captionpos=b,
  keepspaces=true,
  numbers=left,
  numbersep=5pt,
  showspaces=false,
  showstringspaces=false,
  showtabs=false,
  tabsize=2
}
\usepackage{hyperref}

\usepackage[section]{algorithm} 
\usepackage{algpseudocode}  
\algblockdefx[Connector]{Connector}{EndConnector}%
  [1]{\textbf{connector}:#1}%
  {\textbf{end connector}}

\usepackage{cleveref}
  \crefname{part}{Part}{Parts}
  \crefname{chapter}{Chapter}{Chapter}
  \crefname{section}{Section}{Sections}
  \crefname{subsection}{Section}{Sections}
  \crefname{subsubsection}{Section}{Sections}
  \crefname{appendix}{Appendix}{Appendices}
  \crefname{equation}{Equation}{Equations}
  \crefname{figure}{Figure}{Figures}
  \crefname{table}{Table}{Tables}
  \crefname{listing}{Listing}{Listings}
  \crefname{definition}{Definition}{Definitions}
  \crefname{lemma}{Lemma}{Lemmata}
  \crefname{theorem}{Theorem}{Theorems}
  \crefname{corollary}{Corollary}{Corollaries}
  \crefname{remark}{Remark}{Remarks}
  \crefname{algorithm}{Algorithm}{Algorithms}

\makeatletter%
\newcounter{algorithmicH} 
\let\oldalgorithmic\algorithmic
\renewcommand{\algorithmic}{%
  \stepcounter{algorithmicH} 
  \oldalgorithmic} 
\renewcommand{\theHALG@line}{ALG@line.\thealgorithmicH.\arabic{ALG@line}}
\makeatother%

\newcommand{\cgkit}{CG-Kit}

\newcommand{\flashx}{Flash-X}

\journal{}

\begin{document}

\hypersetup{pageanchor=false}
\begin{frontmatter}

\title{\cgkit: \textit{Code Generation Toolkit} for Performant and Maintainable
Variants of Source Code Applied to \flashx\ Hydrodynamics Simulations}

\author[vt,anl]{Johann Rudi\corref{cor1}}
\ead{jrudi@vt.edu}

\author[anl]{Youngjun Lee}
\ead{leey@anl.gov}

\author[vt]{Aidan H. Chadha}
\ead{aidanchadha03@vt.edu}

\author[riken]{Mohamed Wahib}
\ead{mohamed.attia@riken.jp}

\author[uchi]{Klaus Weide}
\ead{kweide@uchicago.edu}

\author[anl]{Jared P. O'Neal}
\ead{joneal@anl.gov}

\author[anl,uchi]{Anshu Dubey}
\ead{adubey@anl.gov}

\cortext[cor1]{Corresponding author}

\affiliation[vt]{
  organization={Department of Mathematics, Virginia Tech},
  addressline={225 Stanger Street},
  city={Blacksburg},
  postcode={24061},
  state={VA},
  country={USA}
}
\affiliation[anl]{
  organization={Mathematics and Computer Science Division, Argonne National Laboratory},
  addressline={9700 S. Cass Avenue},
  city={Lemont},
  postcode={60439},
  state={IL},
  country={USA}
}

\affiliation[riken]{
  organization={RIKEN Center for Computational Science},
  addressline={7-1-26 Minatojima-minami-machi},
  city={Kobe},
  postcode={650-0047},
  country={Japan}
}

\affiliation[uchi]{
  organization={Department of Computer Science, University of Chicago},
  addressline={5730 South Ellis Avenue},
  city={Chicago},
  postcode={60637},
  state={IL},
  country={USA}
}

\hypersetup{
  pdftitle={CG-Kit and Flash-X},
  pdfauthor={Rudi, Lee, et al.}
}

\begin{abstract}

\cgkit\ is a new code generation toolkit that we propose as a solution for
portability and maintainability for scientific computing applications.
The development of \cgkit\ is rooted in the urgent need created by the shifting
landscape of high-performance computing platforms and the algorithmic
complexities of a particular large-scale multiphysics application: \flashx.
This combination leads to unique challenges including handling an existing large
code base in Fortran and/or C/C++, subdivision of code into a great
variety of units supporting a wide range of physics and numerical methods, different
parallelization techniques for distributed- and shared-memory systems and
accelerator devices, and heterogeneity of computing platforms requiring
coexisting variants of parallel algorithms.  All of these challenges demand that
scientific software developers apply existing knowledge about domain applications,
algorithms, and computing platforms to determine custom abstractions
and  granularity for code generation.  There is a critical lack of tools to
tackle these problems.
\cgkit\ is designed to fill this gap with standalone tools that can be combined
into highly specific and, we argue, highly effective portability and
maintainability tool chains.
Here we present the design of our new tools: parametrized source trees,
control flow graphs, and recipes.  The tools are implemented in
Python. Although the tools are agnostic to the programming language of
the source code,
we focus on C/C++ and Fortran.
Code generation experiments demonstrate the generation of variants of parallel
algorithms: first, multithreaded variants of the basic AXPY operation
(scalar-vector addition and vector-vector multiplication) to introduce the
application of \cgkit\ tool chains; and second, variants of parallel algorithms
within a hydrodynamics solver, called Spark, from \flashx\ that operates on
block-structured adaptive meshes.
In summary, code generated by \cgkit\ achieves a reduction by over
60\% of the original C/C++/Fortran source code.

\end{abstract}

\begin{keyword}

Code generation \sep
Performance portability \sep
Algorithm variants \sep
Syntax tree \sep
Control flow graph \sep
Multiphysics simulation

\end{keyword}

\end{frontmatter}
\hypersetup{pageanchor=true}

\section{Introduction}
\label{sec:intro}

Scientific computing at large scales is at the cusp of transformation
in several ways. The longstanding traditional use of
high-performance computing (HPC) resources for simulations
\cite{DubeyWeideONeilEtAl22, RudiMalossiIsaacEtAl15}
is now often just one of many aspects of HPC use in scientific
workflows. Computational and modeling techniques have become more
sophisticated \cite{RudiStadlerGhattas17, RudiShihStadler20, Couch2021},
a trend that is typically accompanied by increase in the
complexity of scientific software.
Scientific workflows face an additional challenge, that of
simultaneously increasing heterogeneity in hardware
architecture. Together these two trends turn the task of writing
efficient and portable scientific software into a formidable one,
unless helpful abstractions and tools are developed to assist in its
design and implementation.

Since the end of Dennard scaling, where higher performance was
achieved through increasing the clock speed of the chips, the trend
has been to obtain greater computing power through massive
parallelism. This has led to the emergence of many-core and GPU-based
architectures, accompanied by C++ template-metaprogramming-based
abstractions~\cite{CarterEdwardsSunderland14,BeckingsaleBurmarkHornungEtAl19,GridTools}
and directives-based solutions~\cite{openmp,openacc}. The former were
helpful in unifying data structures and micro-parallelism that
differed between CPUs and GPUs---assuming a C++ code base and that one
was prepared for more challenges in code and tooling complexity. The latter left more
control in the hands of code developers but was not as efficient
at unifying the code when different data layouts were needed for
different devices.
Neither approach is particularly well suited for unifying code when
differences occur at algorithmic levels.

Several in the community now believe that in the future the
only way to obtain more performance from hardware will be through
specialization, which will require chiplets for specific functions
embedded in the CPU, and a variety of accelerators for different
functionalities needed~\cite{10.1145/3552309}.
In this scenario, control flow is likely to get more
complicated because of the data movement and computation mapping
requirements. We have developed a portability solution with a
collection of tools that aim to ease the job of scientific code
developers in this highly challenging environment. The solution
includes three tools: (1)~a \emph{code generation toolkit, \cgkit},
that equips developers with a collection of modular and composable tools
that can be used to unify high-level algorithmic variants and can
also be used to describe the map of computation to hardware
components; (2)~a runtime data movement tool, \emph{Milhoja},~\cite{ONeilWahibDubeyEtAl22} that
manages the movement of data and computation; and (3) a~\emph{macroprocessor}
\cite{DubeyLeeKlostermanVatai23} that provides the ability to unify computation
at the level of data structures and micro-parallelism similar to C++
abstractions.

The spotlight of this paper is on \cgkit\ and, in particular, on
\cgkit-enabled generation of code variants. The other two tools are
described in the corresponding references. Here, by variants we mean different realizations
of numerical algorithms that lead to the same solution outcome but
differ in the details of algorithm design and/or the implementation
of how the solution is obtained. As mentioned earlier, the need for
variants arises from differences in hardware architecture.
Maintaining all variants explicitly can lead to code bloat that can
make the code hard to maintain. With \cgkit\ the variants can be
expressed succinctly as \emph{\cgkit\ recipes} in the Python
language.  The recipes are translated into \emph{\cgkit\ parameterized
source trees}.  Platform-dependent customizations are enabled by
\emph{\cgkit\ templates} that comprise the building blocks of
parameterized source trees.  Our tools parse source code for any
programming language generally; however, in the context of scientific computing
we focus on the C/C++ and Fortran languages specifically.  The final
code is compilable and optimized for readability by human programmers,
which is a key property to aide developers with code understanding, debugging, and
reasoning about performance metrics.

\begin{figure*}[t]
  \centering
  \includegraphics[width=\textwidth]{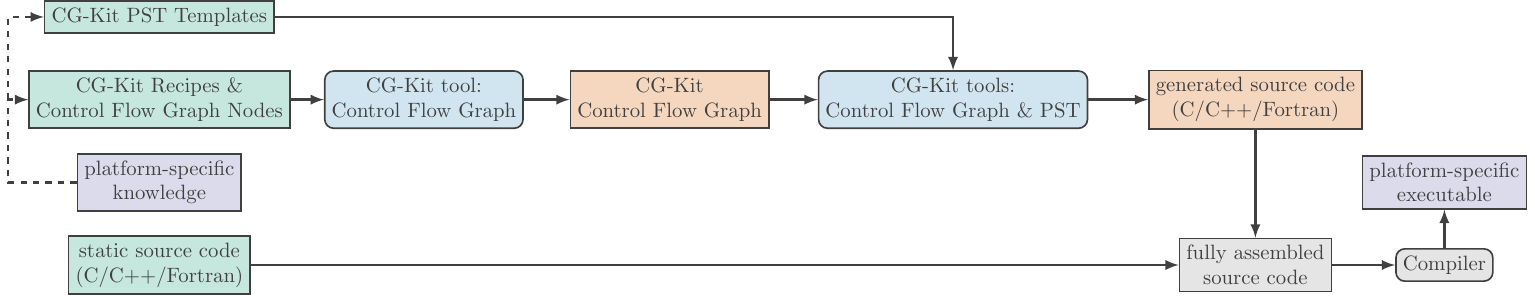}
  \caption{Chain of \cgkit\ tools for code generation when PSTs, control flow
    graphs, and recipes are used.
    Green boxes are user input files; orange boxes represent intermediate
    outputs of tools; blue boxes depict \cgkit\ tools.
    Platform-specific knowledge is provided by the user (left purple box), and
    the compiled program is a platform-specific executable (right purple box).
    Note that the difference to the tool chain in \cref{fig:toolchain-ctree} is
    the addition of the control flow graph.
  }
  \label{fig:toolchain-cflow-ctree}
\end{figure*}

We demonstrate the use of \cgkit\ in \flashx\
\cite{DubeyWeideONeilEtAl22}, a multiphysics simulation software used
by several science domains. It is the new incarnation of
\cite{DubeyAntypasGanapathyEtAl09} designed from the ground up for
portability across a wide variety of platform architectures.
Our portability solution
described above was designed with \flashx\ as its use case,
but our tools are kept general to support any application,
and they can be used in standalone mode or in combination as a tool chain;
see the illustration in \cref{fig:toolchain-cflow-ctree} that will be explained
over the course of \cref{sec:methods}.

\section{Background, Related Work, and Contributions}

Once GPUs became usable for floating-point operations, their adoption
by the HPC scientific community was inevitable given their performance and
energy efficiency advantages. At the same time the challenges posed by
having to move data between devices and the possibility of
computations themselves being different on different devices brought
focus on programming models and abstractions as fundamental needs~\cite{padal2014,Mittal_2015}.
The solutions have
taken several forms that can be broadly categorized into four
types described next.

\subsection{Abstractions and programming models}
\label{sec:abs}

The earliest practical solution for using GPUs was CUDA as a specialized language
supported by Nvidia 
generalizing their GPUs
for scientific work. This was soon followed by directives-based solutions, such
  as,
OpenACC~\cite{openacc} and OpenMP~\cite{openmp}.
The directives
gave fine-grained control of parallelism to the developer and have
largely been incorporated into the compilers directly.

Another early approach toward abstraction was using  domain-specific languages
(DSLs), for instance, \cite{halide,gysi2015stella,clement2018claw,nebo}. Several of these
DSLs had success in becoming a good solution for their target
communities. 
Generally, however, the burden of
growing the DSL with growth of the software proved to be too large
for smaller groups, and some switched over to the third class of
solutions. These are abstractions based on C++ template-metaprogramming,
such as Kokkos~\cite{CarterEdwardsSunderland14},
Raja~\cite{BeckingsaleBurmarkHornungEtAl19},
AMReX~\cite{ZhangAlmgrenBecknerEtAl19}, and
STELLA/GridTools~\cite{GysiOsunaFuhrerEtAl15},
 which enable unification of variants that
arise out of different computational requirements of CPU vs.\ GPU. They
heavily rely on C++ templates to describe the computation, which then
can be generated to the specialized target device, as needed. Some tools
(e.g., Legion~\cite{legion}) also provide
asynchronization of data movement along with compute abstractions.

The final set consists of languages specially designed for HPC
workloads, for example, Chapel~\cite{chapel} or Co-array Fortran~\cite{coarray}.
These languages have struggled to get widespread
adoption, so their future remains in question. Several features of
Co-array Fortran have been incorporated into the Fortran
standard. Several attempts also have been made to leverage the strong
ecosystem of Python, because it has the fastest growing-number of
users. Attempts have been made with varying degrees of success~\cite{dace,BehnelBradshawCitroEtAl11}, the most
notable being the AI/ML frameworks PyTorch~\cite{pytorch} and
TensorFlow~\cite{tensorflow}, which both support linear algebra routines with an
API for operations on array-structured data that is similar or identical to
NumPy~\cite{numpy}.

As scientific applications such as \flashx\ become more adopted, the
complexities of their implementation increase dramatically.
Increasingly heterogeneous architectures---with
CPUs and a variety of 
accelerators 
having different computing throughput and memory bandwidth---require an
intricate balancing act of cost-benefit trade-offs in software
design. Tools that allow design cognizant of these challenges and
trade-offs are a necessity if we are to continue to see gains
in scientific discovery through computing.

Primary insight that fed into the design of our performance portability
solution is that different aspects of performance portability need
different treatments that are orthogonal to one another. Because
computations are distributed across a plethora of devices, the code
design needs three mechanisms: unification of
code variants dictated by different hardware needs, description of mapping
of computation to hardware resources, and a mechanism to understand and
execute the map. 

\subsection{\flashx}
\label{sec:flashx}

\flashx\ has been the motivator for the development of the portability layers
\cite{RudiONeilWahibEtAl21, ONeilWahibDubeyEtAl22, DubeyLeeKlostermanVatai23}
and also its first use case. Therefore, for
completeness, we describe features of \flashx's infrastructure that
are pertinent to the \cgkit\ discussion. \flashx's source code has a modular
architecture where independent components occur at various levels of a code
hierarchy and at different granularities. A component in \flashx\ is a
self-describing entity that carries metadata about how it fits into
the overall application. A scientific application with \flashx\ is put
together by specifying the required components in a specialized
code unit called {\em Simulation}. A Python tool parses the
requirements and starts assembling the components by recursively
following the requirements of various components specified in the
Simulation unit. At the coarsest granularity, we have infrastructure
or physics units that provide solvers for specific physical models, or a distinct
infrastructural functionality. The {\em Grid} unit is an example of an
infrastructure unit that encompasses all the support needed by the
discretized mesh.  Within the highest-level unit, subunits can
exist at arbitrary granularities. The Grid unit, for instance, has several subunits for
different features, such as boundary conditions, generic solvers (e.g.,
multigrid), or support of particles that interact with the
mesh. Among physics units a good example is the {\em Hydro} unit that
includes multiple solvers for compressible hydrodynamics and
magneto-hydrodynamics. Different solvers exist as alternative
implementations of the unit's API through which it interacts with
other units. Similarly the {\em Equation-Of-State (EOS)} unit provides
implementations for a variety of equations of state that may be used by applications
instances in \flashx. At the finest granularity we can have a single
function or even a macro within a function that can be a component in
its own right; see \cite{DubeyLeeKlostermanVatai23} for more details.

\flashx\ has another orthogonal method of apportioning work through
domain decomposition. It uses structured adaptive mesh refinement
(AMR) where the physical domain is divided into blocks of discrete
data points called cells. Different blocks have identical number of
cells, although the physical spacing between cells may vary between
blocks. A block includes a surrounding halo of ghost cells, which
then makes a block a subdomain that is (for implementation
purposes) indistinguishable from the entire domain for a physics unit
operating on it.

\flashx's high level of composability within the code and availability of
 multiple blocks that can be distributed variably among computational
 resources make it possible to  realize an application in many
 different ways when diverse resources
 are available. For instance, it may be possible to overlap certain
 computations with one another if they affect a disjointed set of
 state variables by changing the order of operations. Or it may be
 possible to schedule movement of blocks between devices such that
 the latency of data movement can be hidden. Typically, an experienced
 user  of \flashx\ is expected to have good intuition about the
 possibilities of different ways of orchestrating computation without
 affecting the correctness of the solution. Our performance portability
 solution is targeted at letting such end users dictate what they
 wish done---without ``coding to metal''---and our code transformation and
 generation tools together create a compilable solution desired by the
 user. \cgkit\ sits at the top of this tool chain, where the high-level
 dependencies and concurrency opportunities are expressed in the form
 of recipes, and in cooperation with other tools in the chain it
 facilitates the generation of optimized code. Note that recipes may
 also be expressed for physics units that have multiple ways of
 organizing their computation in different circumstances. In this work
 we use algorithmic variants of {\em Spark}~\cite{Couch2021}, one of the newest hydrodynamics
 solvers, to demonstrate conversion from a recipe to generated code that
 can be compiled. We use Spark as an example because it demonstrates
 the key features of \cgkit\ for code unification in the context of
 variants.
 Going beyond variants, \cgkit\ will play a critical role for code unification
 of end-to-end multiphysics applications using our complementary performance
 portability tools (e.g., \cite{ONeilWahibDubeyEtAl22}) in the future.

\subsection{Contributions}

The present work introduces \cgkit.  We propose a set of standalone tools that can be
combined into highly specific and highly effective portability and
maintainability tool chains. These include \cgkit\ parametrized source
trees, \cgkit\ control flow graphs, and \cgkit\ recipes.
Parametrized source trees (PSTs) allow the expression of structure about source
code using specific knowledge about a scientific application, algorithms, and
computing platform.  Platform-dependent customizations are enabled by PST
templates that constitute the building blocks of PSTs.
Control flow graphs are derived from directed acyclic graphs and represent code
generation operations, such as steps of an algorithm with dependencies between
operations.
Recipes are the user interface to create control flow graphs in a concise manner
using Python.
The most advanced version of a \cgkit\ tool chain that we present here is
an ``end-to-end'' solution:
(i) a recipe creates a control flow graph;
(ii) a control flow graph is traversed to build a PST from PST templates;
(iii) a PST is parsed into (C/C++/Fortran) source code.
\cgkit\ tools enable users to apply their knowledge about domain applications,
algorithms, and computing platforms to customize abstractions and
select a desired granularity for code generation.  The generated code is
optimized for readability by human programmers, because it is key to
debugging application code, clearly connecting input/output relationships for
generated code, and reason about performance of generated code and hence its
portability across platforms.

\section{Code Generation Methods with \cgkit}
\label{sec:methods}

\subsection{Tree-based source code transformation}
\label{sec:ctree}

We propose a new technique that exploits tree topologies as a representation of
source code.  Our approach is based on, first, a simplification of abstract
syntax trees and, second, expression of structure within source code.

\subsubsection{Background: Abstract Syntax Trees}
\label{sec:ast}

Abstract syntax trees (ASTs)~\cite{DaveDave12,Harper16} represent source code
of programming languages, such as C/++, Fortran, and Python, in an abstract
tree structure.  ASTs represent every detail in a programming language.  This
results in rich context information that can be utilized
for code transformation~\cite{CordyHalpern-HamuPromislow91,Baxter02,NeculaMcPeakRahulEtAl02}.
An advantage of AST-based code transformation is the ability to ingest source
code directly without needing intervention from programmers.  On the flip side,
developers face increased complexity to control AST-based code transformations,
because the information-dense ASTs need to be efficiently parsed and managed.  As a result, the
development effort is shifted from modifying the source code to the control of code
generation tools, assuming appropriate tools exist.

One reason for the described shift of programming burden is the lack of a way to
express additional structures within source code that is specifically helpful in code
transformation.  We propose to address this gap with \cgkit's parametrized
source trees (\cref{sec:pst}) and decomposition of source code into templates
(\cref{sec:tpl}).

\subsubsection{\cgkit\ Parametrized Source Trees}
\label{sec:pst}

We propose a new tree structure representation for source code, which we call
\emph{parametrized source trees}, where the tree arises from
expressions entered alongside the code; this is illustrated in
\cref{lst:pst-demo-expr}.  This implies that the PST expressions can annotate
source code of any programming language and PSTs are universally applicable
across programming languages.
Programmers decide about the placement of expressions (e.g.,
\texttt{\_connector:function} and \texttt{\_link:kernel} in
\cref{lst:pst-demo-expr}), therefore intentionally imposing the tree structure
using their domain knowledge about the code as well as their desired granularity
of code transformation.  The expression of structure about the source code
allows code transformation to be controlled with less implementation
complexity using \cgkit's complementary PST-based tools (see
\cref{sec:ctrflow}).  Furthermore, PSTs naturally enable the decomposition of
code into \cgkit\ PST templates (see \cref{sec:tpl}).

\begin{figure}[tb]
  \def\lstLabel{lst:pst-demo-expr}
  \def\lstCaption{Illustration of PST expressions placed in snippets of C source
  code. The listing shows the content of two files: one containing a function
  (top) and the other a computational kernel (bottom). The expressions
  \texttt{\_connector:ID} and \texttt{\_link:ID} are PST annotations, where
  \texttt{ID} refers to user-defined identifiers.}
  \input{inc/lst_pst_demo_expressions}
\end{figure}

\begin{figure}[tb]
  \centering
  \includegraphics[width=0.7\columnwidth]{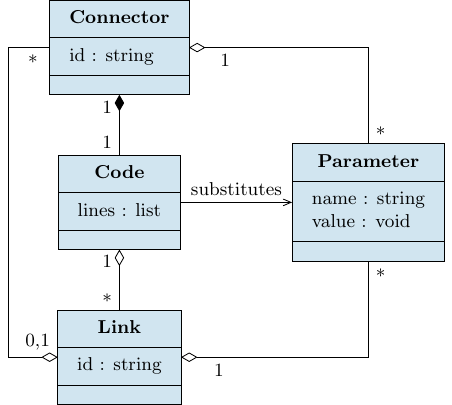}
  \caption{UML diagram of a parametrized source tree (PST).
  A PST's root is a connector, which comprises one instance of code. The
  lines of code contain an arbitrary number of links.  Any number of
  connectors, in turn, can attach to links, and they are matching by unique id's.
  This cycle is how more layers in the tree structure are built.
  Parameters are defined for connectors and for links, and they are substituted
  by referring to them within lines of code.}
  \label{fig:PST}
\end{figure}

The design for a PST is given in \cref{fig:PST} using the class diagram in the
Unified Modeling Language (UML).  \cref{fig:PST} shows that a PST is based on
just a few simple components:
\begin{enumerate}
  \item source code,
  \item parameters,
  \item links, and
  \item connectors.
\end{enumerate}
The class Connector represents also the root of a PST, and it is composed of the
source code (class Code).  The lines within the class Code contain regular
source code and any number of links (class Link).  Any of these links, in
turn, provides locations where connectors can attach, hence allowing the PST to
grow with one tree level being represented by connector--code--link.
The class Parameter allows the hierarchy of connector--code--link levels of the
PST to propagate context information between the levels.  A parameter (i.e., a
name-value pair) is passed down the tree hierarchy such that parameter
definitions from upper levels can be used at lower levels of the tree.
To access a parameter's value within class Code, one simply refers to
it by its (unique) name.  A substitution of the parameter name with its value
takes place when a PST is parsed.
Based on the previous illustration in \cref{lst:pst-demo-expr}, we now
incorporate parameters as additional annotations in the code; see
\cref{lst:pst-demo-params}.

\begin{figure}[tb]
  \def\lstLabel{lst:pst-demo-params}
  \def\lstCaption{Illustration of PST parameters, continuing the illustration in
  \cref{lst:pst-demo-expr}. Parameters are defined as \texttt{\_param:NAME =
  VALUE}, where \texttt{NAME} and \texttt{VALUE} are provided by the user.  The
  primary use for parameters is to propagate information between levels of the
  PST. Here, the variables \texttt{a}, \texttt{x[i]}, and \texttt{y[i]} are
  propagated from the function to the kernel.}
  \input{inc/lst_pst_demo_parameters}
\end{figure}

While PSTs follow a simple design, they are not limited in versatility,
and the decision about the precise levels of the PST is given to programmer, who
can utilize domain and platform knowledge.

It is easy to verify correctness of a PST and also easy to assert the existence
of a definition for all parameters that are required from the source code at each
tree level.  The main benefit of PSTs, however, is that they are easily
inspectable by humans, because PSTs are internally represented as Python
dictionaries and, as such, can be directly output in an accessible JSON format.
To allow for further tracing of PSTs once they are parsed, a verbosity flag
injects additional commented lines in the output source code that show
connectors and links as well as the templates that a PST was composed of.
These two output features, JSON and tracing in parsed code, support users in
reasoning about which inputs are responsible for which outputs.
We continue the illustration of \cref{lst:pst-demo-params} and present a JSON
output of the PST upon connecting the code from file \texttt{kernel.c} to the
link inside file \texttt{function.c}; see \cref{lst:pst-demo-json}.
The corresponding code parsed from that PST with activated verbosity is given in
\cref{lst:pst-demo-parsed}.

\begin{figure}[tb]
  \def\lstLabel{lst:pst-demo-json}
  \def\lstCaption{Illustration of a JSON output of a PST, continuing
  the illustration in \cref{lst:pst-demo-params}.
  In the shown PST the kernel connector from file \texttt{kernel.c} is attached
  to the matching link in file \texttt{function.c} (both shown in
  \cref{lst:pst-demo-params}).}
  \input{inc/lst_pst_demo_json}
  \def\lstLabel{lst:pst-demo-parsed}
  \def\lstCaption{Illustration of parsed code from the PST shown in
  \cref{lst:pst-demo-json} with activated verbosity.  The printed tags allow the
  user to trace which input files caused the final output.}
  \input{inc/lst_pst_demo_parsed}
\end{figure}

\subsubsection{\cgkit\ PST Templates}
\label{sec:tpl}

Along with PSTs, we have implicitly introduced \emph{PST templates} in the
preceding \cref{sec:pst}, and we have provided examples in
\cref{lst:pst-demo-expr,lst:pst-demo-params}.  Templates constitute the building
blocks of PSTs.  A template is defined as a file as follows:
\begin{enumerate}
  \item it must contain a single or multiple connectors (by stating
    \texttt{\_connector:ID} with some user-defined \texttt{ID});
  \item a connector is followed by lines of source code;
  \item none or multiple links can be placed within lines of code (by stating
    \texttt{\_link:ID} with another user-defined \texttt{ID}).
\end{enumerate}
One such template represents one level of the PST consisting of
connector--code--link.  The PST is extended in depth by adding additional
connector--code--link levels, which are coming from other templates and attached
to the existing tree.  The connectors of a to-be-included template are matched
to the existing links in the PST.  Therefore, a requirement for the feasibility
of adding a particular template is that all of the connectors have id's that
match the id's of links of the PST.

The hierarchical structure of a PST requires that information can be passed
between different levels of the tree.  This is addressed by defining parameters inside of
templates.  The parameters can be used at any level of the tree below their
definition; hence, parameters are the means to propagate information toward the
leaves of the tree.
Template files that are written with the intent to be included at a lower level
(e.g., the template in file \texttt{kernel.c} in \cref{lst:pst-demo-params})
uses parameters that are expected to be defined at a higher level in the PST.

The composition of different templates into different PSTs is what enables
the generation of variants of source code, because different templates can be
selected to extend a PST as long as the included templates' connectors are
compliant with the existing links of the PST.
The design of PSTs intentionally leaves the granularity of the decomposition of
code into templates entirely up to users.
It allows users to adapt the implementation of code generation to the needs of a
numerical method or algorithm and to the platforms that they aim to support.

\begin{figure*}[t]
  \centering
  \includegraphics[width=0.66\textwidth]{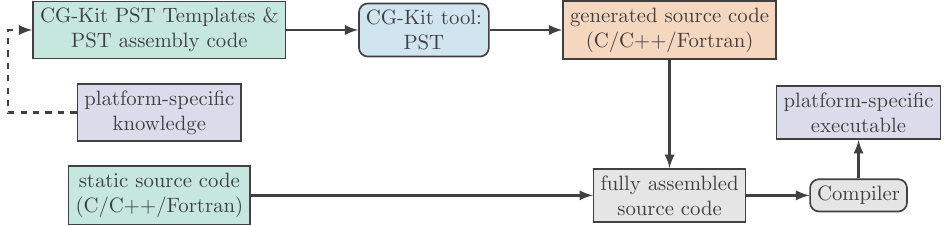}
  \caption{Chain of \cgkit\ tools for code generation when only PSTs are used.
    Green boxes are user input files; orange boxes represent intermediate
    outputs of tools; blue boxes depict \cgkit\ tools.
    Platform-specific knowledge is provided by the user (left purple box), and
    the compiled program is a platform-specific executable (right purple box).
  }
  \label{fig:toolchain-ctree}
\end{figure*}

In summary, \cgkit\ PSTs and templates can be employed by users to generate
variants of algorithms and, therefore, platform-specific code. We illustrate
such a tool chain in \cref{fig:toolchain-ctree}.
The boxes on the left column of the figure denote the user input files and
knowledge; \cgkit\ PST is a tool that generates code from platform-specific
templates, which, after combining with a static (i.e., non-generated and
platform-independent) code, can be compiled into a platform-dependent
executable.

\subsection{Graph-based control flow description}
\label{sec:ctrflow}

The code generation based on PSTs, proposed in \cref{sec:ctree}, can be
directly utilized by users (see the tool chain in \cref{fig:toolchain-ctree}).
Additionally, we propose PST-based tools that automate code generation workflows
generally and target generating variants of codes specifically.

In this section we first identify patterns for code generation, from which we
subsequently derive the \cgkit\ recipe interface, which generates a \cgkit\
control flow graph.  The---typically concise---recipes represent an abstraction
of an algorithm into \emph{code generation operations} and, as such, can be used
for possible variants.  \cgkit\ maps a recipe to a control flow graph, which
then enables the construction of PST-based source code from an implemented
recipe.

\subsubsection{Taxonomy of Patterns}
\label{sec:patterns}

This section identifies several patterns that we aim to support for code
generation with \cgkit, where we focus on only the patterns relevant for code
generation of variants.
The main concepts for the subsequently introduced patterns are
\begin{enumerate}
  \item stream of code generation operations (e.g., a step of an algorithm
    applied on one data item),
  \item dependencies between operations, and
  \item concurrency of data items.
\end{enumerate}
While describing the patterns, we will illustrate them with graphs as well as
recipes, where the latter are formally introduced later in \cref{sec:recipe}.

\paragraph{Pattern: Pipeline}

\begin{figure}[tb]
  \centering
  \tikzset{
  actionSty/.style={circle, draw, thick, darkgray, fill=backcolor, minimum size=2.0em, align=center},
  edgeSty/.style={->, >=latex, very thick, darkgray}
}
\begin{tikzpicture}
  \def\d{2.0}

  \node[actionSty] (R) at (1*\d,0) {$R$};
  \node[actionSty] (S) at (2*\d,0) {$S$};
  \node[actionSty] (T) at (3*\d,0) {$T$};
  \node[actionSty] (X) at (2.5*\d,-0.7*\d) {$X$};
  \node[actionSty] (Y) at (3.5*\d,-0.7*\d) {$Y$};
  \node[actionSty] (Z) at (4*\d,0) {$Z$};

  \path (R) edge[edgeSty] (S);
  \path (S) edge[edgeSty] (T);
  \path (S) edge[edgeSty] (X);
  \path (X) edge[edgeSty] (Y);
  \path (Y) edge[edgeSty] (Z);
  \path (T) edge[edgeSty] (Z);
\end{tikzpicture}
  \caption{Illustration of the pipeline pattern as a graph with the nodes being
    code generation operations (e.g., $R$, $S$, $T$, \ldots) and the arrows
    indicating dependencies (e.g., $S$ depends on completion of $R$).}
  \label{fig:pattern-pipeline}
  \vskip 2ex
  \def\lstLabel{lst:pattern-pipeline}
  \def\lstCaption{Illustration of the pipeline pattern, corresponding to the
    graph in \cref{fig:pattern-pipeline}, in a \cgkit\ recipe.}
  \input{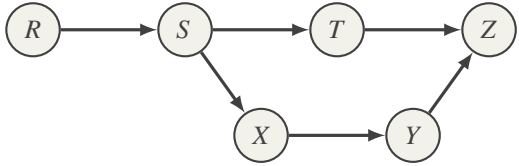}
\end{figure}

The first pattern is fundamental for the realization of \cgkit\ control flow
graphs that are presented in \cref{sec:graph}.
A pipeline expresses the execution order of code generation operations and the
dependencies of operations on one another.
If operations are meant to be applied to data items (e.g., discretized
spatial/temporal operators of a partial differential equation), then those
data items would flow through the pipeline concurrently; hence, the data items
are assumed to be independent of each other.
We illustrate the pipeline pattern with the graph in \cref{fig:pattern-pipeline}
and list the corresponding \cgkit\ recipe for that graph in
\cref{lst:pattern-pipeline}.
Note that in \cref{lst:pattern-pipeline} the variables on the left-hand side of
the equal sign are handles, which are used to indicate dependencies between
code generation operations.  In particular, they do not represent output data
generated by an operation.

\paragraph{Pattern: Begin-End}

\begin{figure}[tb]
  \centering
  \tikzset{
  actionSty/.style={circle, draw, thick, darkgray, fill=backcolor, minimum size=2.0em, align=center},
  attrSty/.style={minimum size=2.0em, align=center, font=\footnotesize},
  opSty/.style={draw, thick, darkgray, fill=jr@lightgreen, align=center, rounded corners=1ex, minimum height=4ex},
  edgeSty/.style={->, >=latex, very thick, darkgray},
  assocEdgeSty/.style={-, semithick, darkgray},
  assocLabelSty/.style={align=center}
}
\begin{tikzpicture}
  \def\d{2.0}

  \node[actionSty] (R) at (0*\d,0) {$R$};
  \node[actionSty] (S) at (1*\d,0) {$S$};
  \node[actionSty] (T) at (2.35*\d,0) {$T$};

  \node[opSty]     (loopBegin) at (1*\d,-0.7*\d)   {LoopBegin};
  \node[actionSty] (X)         at (2*\d,-0.7*\d)   {$X$};
  \node[actionSty] (Y)         at (2.7*\d,-0.7*\d) {$Y$};
  \node[opSty]     (loopEnd)   at (3.6*\d,-0.7*\d) {LoopEnd};
  \node[assocLabelSty] (knows) at (2.3*\d,-1.2*\d) {knows};

  \node[actionSty] (Z) at (3.6*\d,0) {$Z$};

  \path (R)         edge[edgeSty] (S);
  \path (S)         edge[edgeSty] (T);
  \path (S)         edge[edgeSty] (loopBegin);

  \path (loopBegin) edge[edgeSty] (X);
  \path (X)         edge[edgeSty] (Y);
  \path (Y)         edge[edgeSty] (loopEnd);

  \path (loopEnd)   edge[edgeSty] (Z);
  \path (T)         edge[edgeSty] (Z);

  \draw[assocEdgeSty] (loopBegin) |- (knows) -| (loopEnd);
\end{tikzpicture}
  \caption{Illustration of the begin-end pattern as a graph with the green
    nodes representing a pair for the beginning and end of a loop, while the
    nodes ($X$, $Y$) are operations performed within the loop.}
  \label{fig:pattern-begin-end}
  \vskip 2ex
  \def\lstLabel{lst:pattern-begin-end}
  \def\lstCaption{Illustration of the begin-end pattern, corresponding to the
    graph in \cref{fig:pattern-begin-end}, in a \cgkit\ recipe.}
  \input{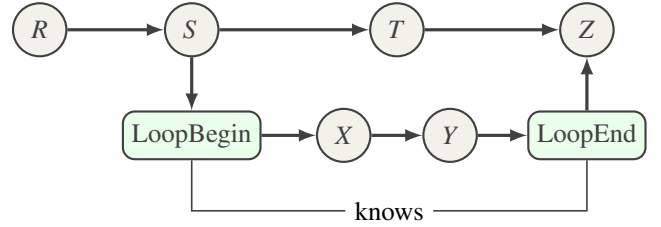}
\end{figure}

A begin-end pattern describes the nesting of (a pipeline of) code generation
operations within a construct that has a defined beginning and an end, for
example, a loop.  The coupling of two nodes in the graph with this pattern
enables the generation of a PST to be performed as nodes of a graph are
visited in the order of their occurrence.
An illustration of a graph containing begin-end nodes is in
\cref{fig:pattern-begin-end} with a corresponding \cgkit\ recipe in
\cref{lst:pattern-begin-end}.  The coupling of the LoopBegin and LoopEnd nodes
ensures that some user-defined initialization tasks can be performed upon visit
of LoopBegin and some finalization tasks can be performed upon visit of
LoopEnd.

\paragraph{Pattern: Concurrent Data}

The concurrent data pattern describes a single operation or a pipeline
of operations executed on independent data items.
This pattern is derived from the begin-end pattern; thus, implementing
the concurrent data pattern involves a pair of begin and end nodes.
The relevance of having this pattern is to enable expressing data parallelism in
a code generation workflow.

\subsubsection{\cgkit\ Recipes}
\label{sec:recipe}

\emph{\cgkit\ recipes} are written by users in the Python language.  They
provide an interface to realize the patterns from \cref{sec:patterns}, and they
create a resulting control flow graph described later in \cref{sec:graph}.
The motivation for recipes is to enable users to abstract building blocks
of algorithms to a desired level such that variants can be easily composed
and modified.  The particular level or degree at which an algorithm's
implementation is abstracted will strongly depend on the algorithm itself.
Therefore, \cgkit\ recipes are not making assumptions about the abstraction.
They are a tool to define the level of abstraction desired by users.
Recipes follow a ``define and run'' principle.  This principle
entails that a recipe defines algorithmic building blocks (e.g., subroutines,
actions, etc.) and their dependencies on each other.

To write a recipe, we begin by creating an instance of the class
\texttt{ControlFlowGraph} (for details see \cref{sec:graph}), for example, in
line 1 of \cref{lst:pattern-pipeline}.
The graph is populated with nodes that are instantiations of user-defined
classes.  Examples of nodes that we presented in \cref{lst:pattern-begin-end}
are from classes \texttt{CodeGenNode}, \texttt{LoopBeginNode}, and
\texttt{LoopEndNode}.
To add a node object to the graph, we utilize a method, \texttt{add}, that has a
\emph{functional syntax} that enables creation of directed edges between nodes
of the graph.  This syntax has regular function arguments to initialize the
node object, and it provides a second pair of brackets that hold the dependency
information.  Valid inputs for the dependency are the handles returned from
calling the \texttt{add} method of the graph or a Python list of these handles.

The result after executing a recipe is a graph, which is described in the
\cref{sec:graph}.

\subsubsection{\cgkit\ Control Flow Graphs}
\label{sec:graph}

Given a recipe from \cref{sec:recipe}, \cgkit\ generates a corresponding
\emph{control flow graph}, which is the recipe represented as a directed
acyclic graph (DAG).
Internally, \cgkit\ control flow graphs leverage the implementation of DAGs from
the Python package NetworkX~\cite{HagbergSchultSwart11}, which provides data
structures for graphs and graph algorithms.
The nodes of the graph represent user-defined code generation operations (e.g.,
steps of an algorithm applied to data items), and the graph's edges represent
dependencies between nodes, such as the order of operations.
Our requirements for a valid control flow graph are as follows:
\begin{enumerate}
  \item the graph is directed and acyclic;
  \item it has a unique root node, $R$, and a unique leaf node, $L$;
  \item any path that starts at $R$ must end at $L$.
\end{enumerate}
As a consequence of the requirements, we can assume that
(i) control flow graphs can be referred to by tuples of root and leaf nodes,
$(R,L)$;
(ii) the longest path of the graph starts at $R$ and ends at $L$; and
(iii) for any node $U$ of the graph, there exists a path from $R$ to $L$ that
visits $U$.

We traverse a control flow graph in a particular way.
We start at the root node and step along directed edges to adjacent nodes.
Prior to being able to step along an outgoing edge of a node, all incoming edges
must have been traversed.  In other words, nodes that have multiple incoming
edges are blocking for the purpose of the traversal.
This traversal protocol ensures that every node of the graph is visited in a way
that is controlled for executing code generation operations, which, in
the current work, are creating and extending a PST with more and more levels.
The execution of (user-defined) operations at nodes is what integrates \cgkit\
PSTs with control flow graphs and, ultimately, with recipes.

A chain of \cgkit\ tools that utilizes recipes, control flow graphs, and PSTs is
illustrated in \cref{fig:toolchain-cflow-ctree}.  This figure shows an extension of
the tool chain from \cref{fig:toolchain-ctree}, where additional input files,
namely, recipes and definitions of nodes for the control flow graph, are provided
by users.  Furthermore, recipes are processed by one \cgkit\ tool for graphs,
which can subsequently utilize another \cgkit\ tool for PSTs, while both tools
can also be used alone.  The decomposition of the code generation workflow
into recipes and templates and using two different tools for their processing
allow \cgkit\ to be modular and address different (likely orthogonal) aspects
of creating platform-specific implementations of algorithms.

\section{Code Generation Experiments}
\label{sec:experiments}

We perform two sets of experiments. The first is an illustration of the usage of
the two \cgkit\ tool chains, shown in \cref{fig:toolchain-ctree} and in
\cref{fig:toolchain-cflow-ctree}, to generate variants of the AXPY operation
from numerical linear algebra.  The second set of experiments comprises variants for
hydrodynamics simulations in \flashx.

\subsection{Illustration of variants for scalar-vector multiplication and
vector-vector addition (AXPY)}
\label{sec:axpy-variants}

We illustrate the usage of our code generation tools with a simple example from
numerical linear algebra: the AXPY operation.  AXPY ($A$ times $X$ plus $Y$) is
a scalar-vector multiplication followed by a vector-vector addition: $y_i = a
x_i + y_i$, where $a\in\mathbb{R}$, $x,y\in\mathbb{R}^N$, $N\in\mathbb{N}$, and
subscript notation $x_i,y_i$ refers to entries of the vectors $x,y$.

In this section the aim is to generate five variants of AXPY, summarized in
\cref{tab:axpy-variants}, which differ in their multithreaded parallel
algorithms and in the implementation of parallelization using either
OpenMP or CUDA.  The C/C++ language is used for all variants of AXPY.

\begin{table}[b]
  \caption{Overview of variants of AXPY operation.}
  \label{tab:axpy-variants}
  \centering
  \footnotesize
  \begin{tabular}{ccc}
    \toprule
    \thead{Variant} & \thead{Multithread Algorithm} & \thead{Parallelization} \\
    \midrule
    1       & \multirow{2}{*}{Increment by 1 (\cref{alg:axpy-incr-1})} & OpenMP \\
    2       &                                                          & CUDA \\
    \midrule
    3       & \multirow{2}{*}{Increment by \#threads (\cref{alg:axpy-incr-threads})} & OpenMP \\
    4       &                                                                        & CUDA \\
    \midrule
    5       & Single iteration (\cref{alg:axpy-single-iter}) & CUDA \\
    \bottomrule
  \end{tabular}
\end{table}

The three different
\cref{alg:axpy-incr-1,alg:axpy-incr-threads,alg:axpy-single-iter}
show a multithreaded function for the AXPY computation that varies in how the
assignment of OpenMP or CUDA threads, $t$, to array entries, $i$, is carried
out.
The AXPY functions of the first two \cref{alg:axpy-incr-1,alg:axpy-incr-threads}
can be used with both OpenMP and CUDA.\footnote{%
  For OpenMP, setting the thread index and the number of threads is done via
  $t\gets\text{\texttt{omp\_get\_thread\_num()}}$ and
  $T\gets\text{\texttt{omp\_get\_num\_threads()}}$, respectively.
  For CUDA, setting the thread index and the number of threads is done via
  $t\gets\text{\texttt{blockDim.x * blockIdx.x + threadIdx.x}}$ and
  $T\gets\text{\texttt{gridDim.x * blockDim.x}}$, respectively.
}
\cref{alg:axpy-incr-1} uses the thread index, $t$, and the number of threads,
$T$, to subdivide the array entries into equally sized blocks with consecutive
indices $i_\mathrm{lo} \le i < i_\mathrm{hi}$; this leads to consecutive memory
access per thread (i.e., intrathread consecutive access).
\cref{alg:axpy-incr-threads}, on the other hand, sets the starting
array index to the thread index, $i_\mathrm{lo}=t$, and iterates through the
arrays with a stride equal to the number of threads, $T$; this leads to
nonconsecutive memory access per thread but consecutive access for a group of
threads (i.e., interthread consecutive access).
\begin{algorithm}[tb]
  \caption{AXPY, increment by 1, OpenMP/CUDA}
  \label{alg:axpy-incr-1}
  \begin{algorithmic}[1]
  %
  \Function{axpy\_increment\_1}{$N,a,x,y$}
    \State $t \gets$ thread index
    \State $T \gets$ number of threads
    \State $i_\mathrm{lo} = \lfloor (N t)/T \rfloor$
    \Comment $\lfloor\cdot\rfloor$ is floor (i.e., integer division)
    \State $i_\mathrm{hi} = \lfloor (N (t+1))/T \rfloor$
    \For {$i=i_\mathrm{lo}$; $i<i_\mathrm{hi}$; $i=i+1$}
      \State $y_i = a x_i + y_i$
    \EndFor
  \EndFunction
\end{algorithmic}

\end{algorithm}
\begin{algorithm}[tb]
  \caption{AXPY, increment by \#threads, OpenMP/CUDA}
  \label{alg:axpy-incr-threads}
  \begin{algorithmic}[1]
  %
  \Function{axpy\_increment\_threads}{$N,a,x,y$}
    \State $t \gets$ thread index
    \State $T \gets$ number of threads
    \For {$i=t$; $i<N$; $i=i+T$}
      \State $y_i = a x_i + y_i$
    \EndFor
  \EndFunction
\end{algorithmic}

\end{algorithm}
\cref{alg:axpy-single-iter} can typically be executed only with CUDA, because
its AXPY function operates only on a single entry of arrays $x,y$ with index
$i=t$.  In turn, the number of threads has to satisfy $T \ge N$, which in CUDA
is done by simply scaling the number of thread blocks relative to the problem
size, $N$.  This variant of AXPY results in intrathread consecutive memory
access similar to \cref{alg:axpy-incr-threads}.
\begin{algorithm}[tb]
  \caption{AXPY, single iteration, CUDA}
  \label{alg:axpy-single-iter}
  \begin{algorithmic}[1]
  %
  \Function{axpy\_single\_iter}{$N,a,x,y$}
    \State $i=t \gets$ thread index
    \If {$i<N$}
      \State $y_i = a x_i + y_i$
    \EndIf
  \EndFunction
\end{algorithmic}

\end{algorithm}

We are using well-known algorithms for AXPY to illustrate our new concepts for
code generation of variants.  This is done in the following
\cref{sec:axpy-variants-ctree,sec:axpy-variants-cflow}, where the former of
the two sections considers code generation using only \cgkit\ PSTs, as depicted
by the tool chain in \cref{fig:toolchain-ctree}, and the latter section
demonstrates code generation using \cgkit\ recipes, control flow graphs, and
PSTs, as depicted by the tool chain in \cref{fig:toolchain-cflow-ctree}.

\subsubsection{Generation of variants with PSTs only}
\label{sec:axpy-variants-ctree}

Code generation exclusively with \cgkit\ PST templates is facilitated by
hierarchically extending a PST using the following tree levels, denoted by
$\ell$,
\begin{itemize}
  \raggedright
  \item $\ell=0$: initial level with driver including \texttt{main} function,
  \item $\ell=1$: variant-specific AXPY implementation and its multithreaded
    execution, and
  \item $\ell=2$: computational kernel, $y_i = a x_i + y_i$.
\end{itemize}
The PST templates at levels $\ell=0,2$ are shared among all variants.
Variant-specific implementations are made by creating one PST template per
variant at level $\ell=1$. Adding the templates files for all the levels and all
the variants amounts to seven files in total.
The level-0 template for the driver is presented (in a concise form) in
\cref{alg:axpy-pst-driver}.  In this algorithm we emphasize the placement of
PST links, which are the positions in the code that allow the PST to be extended
by an additional level.
\begin{algorithm}[tb]
  \caption{PST template at $\ell=0$ for driver (all variants)}
  \label{alg:axpy-pst-driver}
  \begin{algorithmic}[1]
  \Connector{driver}
    \Statex \textbf{parameters}:
    \{$N\gets$ \texttt{length},
      $a\gets$ \texttt{1.0},
      $x\gets$ \texttt{h\_x},
      $y\gets$ \texttt{h\_y},
      $k\gets$ \texttt{2}\}
    \State Include header files
    \State \textbf{link}:include
    \State \textbf{link}:function
    \Function{main}{}
      \State \textbf{link}:variables
      \State Allocate arrays $x,y\in\mathbb{R}^N$
      \State Initialize entries in $x,y$
      \State \textbf{link}:setup
      \State \textbf{link}:execute
      \State Calculate \& print max error over all entries of $y$
      \State \textbf{link}:clean
      \State Deallocate arrays $x,y$
    \EndFunction
  \EndConnector
\end{algorithmic}

\end{algorithm}
The matching connectors for the corresponding links of the driver are present in
every variant-specific level-1 template.  We summarize the presentation of
these five level-1 templates by showing one template for OpenMP in
\cref{alg:axpy-pst-openmp} and one for CUDA in \cref{alg:axpy-pst-cuda}.  In
practice, there exist two templates for OpenMP implementing AXPY
\cref{alg:axpy-incr-1,alg:axpy-incr-threads}, respectively; and there exist three
templates for CUDA implementing AXPY
\cref{alg:axpy-incr-1,alg:axpy-incr-threads,alg:axpy-single-iter}, respectively.
\begin{algorithm}[tb]
  \caption{PST template for OpenMP (variants no.\ 1, 3)}
  \label{alg:axpy-pst-openmp}
  \begin{algorithmic}[1]
  \Connector{include}
    \State Include OpenMP header file
  \EndConnector
  \Connector{function}
    \Statex \textbf{parameters}:
    \{$a\gets$ \texttt{a},
      $x_i\gets$ \texttt{x[i]},
      $y_i\gets$ \texttt{y[i]}\}
    \State Implementation of AXPY function
    \Statex \Comment one variant from \cref{alg:axpy-incr-1,alg:axpy-incr-threads}
  \EndConnector
  \Connector{variables}
    \State Create variables for \#threads and elapsed time
  \EndConnector
  \Connector{setup}
    \State Get number of OpenMP threads
  \EndConnector
  \Connector{execute}
    \State Call AXPY function in parallel
    \Comment warm-up run
    \State Begin timing
    \State Repeat $k$ times: Call AXPY function in parallel
    \State End timing
    \State Print \#threads \& elapsed time
  \EndConnector
  \Connector{clean}
    \Comment no op.
  \EndConnector
\end{algorithmic}

\end{algorithm}
\begin{algorithm}[tb]
  \caption{PST template for CUDA (variants no.\ 2, 4, 5)}
  \label{alg:axpy-pst-cuda}
  \begin{algorithmic}[1]
  \Connector{include}
    \State Include CUDA header file
  \EndConnector
  \Connector{function}
    \Statex \textbf{parameters}:
    \{$a\gets$ \texttt{a},
      $x_i\gets$ \texttt{x[i]},
      $y_i\gets$ \texttt{y[i]}\}
    \State Implementation of AXPY kernel function
    \Statex \Comment one variant from
      \cref{alg:axpy-incr-1,alg:axpy-incr-threads,alg:axpy-single-iter}
  \EndConnector
  \Connector{variables}
    \State Create variables for \#threads, elapsed time, and device memory
  \EndConnector
  \Connector{setup}
    \State Set number of threads per block and number of blocks
    \Statex \Comment different for Algo's \ref{alg:axpy-incr-1},
      \ref{alg:axpy-incr-threads} and Algo.\ \ref{alg:axpy-single-iter}
    \State Allocate arrays $x^{(d)},y^{(d)}\in\mathbb{R}^N$ in device memory
  \EndConnector
  \Connector{execute}
    \State Memcopy arrays from host $\{x,y\}$ to device $\{x^{(d)},y^{(d)}\}$
    \State Launch AXPY kernel on device
    \Comment warm-up run
    \State Begin timing
    \State Repeat $k$ times: Launch AXPY kernel on device
    \State End timing
    \State Memcopy array from device $\{y^{(d)}\}$ to host $\{y\}$
    \State Print \#threads \& elapsed time
  \EndConnector
  \Connector{clean}
    \State Deallocate arrays $x^{(d)},y^{(d)}$ in device memory
  \EndConnector
\end{algorithmic}

\end{algorithm}
The level-2 template of the computational kernel is the same for all
variants, and we already presented this PST template at the bottom of
\cref{lst:pst-demo-params} in \cref{sec:pst}.  Because the arithmetic operation
is the same across the variants, it can be isolated into its own PST template
and reused.

The templates are designed such that each template's C/C++ source code is
independent of the others'.  The dependencies are instead encoded in \cgkit's
PST syntax (i.e., links, connectors, and parameters). Therefore the C/C++ source
code can be treated ``orthogonally'' to the management of code generation with
PST templates.
To demonstrate our choice of PST parameters, we show, for brevity, a subset of
the parameters in
\cref{alg:axpy-pst-driver,alg:axpy-pst-openmp,alg:axpy-pst-cuda}, which are
stated below the beginning of a connector.  In the driver
\cref{alg:axpy-pst-driver}, for instance, the parameters $\{N,a,x,y,k\}$ are
chosen because these are variables in the C/C++ source code that propagate to
the subsequent level-1 templates shown in
\cref{alg:axpy-pst-openmp,alg:axpy-pst-cuda}.  Therefore, defining these
parameters in the driver template ensures that consistent variable names are
used in the level-1 templates.  Recall from \cref{sec:pst} that this is ensured
because parameters in a PST propagate from higher to lower levels of the tree.

The C/C++ code that is generated from PSTs is human-readable---including
consistent indentation---and, in fact, the code is indistinguishable from
regular (i.e., nongenerated) codes.  Hence, debugging output of \cgkit's tool
chain is straigthforward, and the compilation can easily be made to complete
without errors or warning messages from the compiler.  Furthermore, each of the
five compiled programs, corresponding to the five AXPY variants, runs without
arithmetic errors.

While the generation of AXPY variants, including driver code, is a simplified
use case for code generation, it is worthwhile to document the amount of code
savings quantitatively.  This is done in the first row of
\cref{tab:axpy-code-reduc} in \cref{sec:axpy-variants-cflow} by counting the
lines of C/C++ code in all of the templates (shown in the table's column
``input'') and comparing this number with the lines of generated code (shown in
the table's column ``generated'').  The metric of relative \emph{C/C++ code
reduction} is calculated as one minus the ratio of input code and generated
code; this metric shows that a reduction of around 40\,\% is achieved.
Next, we extend the \cgkit\ tool chain to recipes and control flow graphs, which
will result in significant additional code reductions.

\subsubsection{Generation of variants with recipes, graphs, and PSTs}
\label{sec:axpy-variants-cflow}

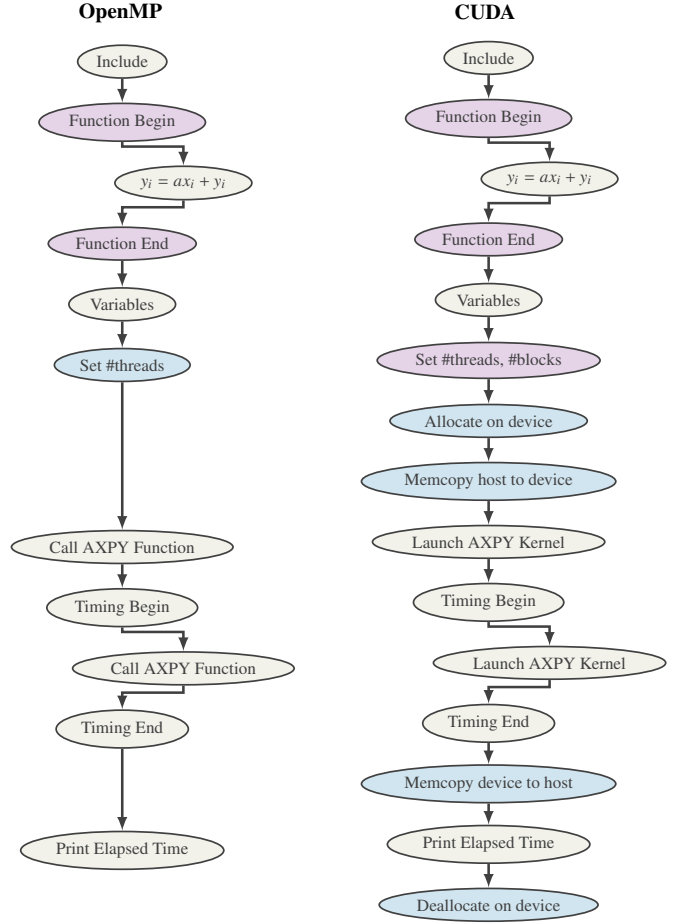
\begin{figure}[tb]
  \footnotesize
  \begin{minipage}[t]{0.448\columnwidth}\vspace{0pt} 
    \hspace{3.6em}\textbf{OpenMP}\\[2ex]
    \resizebox{\columnwidth}{!}{ \tikzset{
  itemSty/.style={ellipse, draw, thick, darkgray, fill=backcolor,
    minimum width=4em, minimum height=2ex, align=center, font=\footnotesize},
  itemH1Sty/.style={itemSty, fill=jr@purple!25!white},
  itemH2Sty/.style={itemSty, fill=jr@lightblue},
  opSty/.style={draw, thick, darkgray, fill=jr@lightgreen,
    minimum height=4ex, rounded corners=1ex, align=center},
  edgeSty/.style={->, >=latex, very thick, darkgray}
}
\begin{tikzpicture}
  \def\d{1.0}

  \node[itemSty  ] (I)  at (0   ,- 0*\d) {Include};

  \node[itemH1Sty] (FB) at (0   ,- 1*\d) {Function Begin};
  \node[itemSty  ] (FK) at (1*\d,- 2*\d) {$y_i = a x_i + y_i$};
  \node[itemH1Sty] (FE) at (0   ,- 3*\d) {Function End};

  \node[itemSty  ] (SV) at (0   ,- 4*\d) {Variables};
  \node[itemH2Sty] (ST) at (0   ,- 5*\d) {Set \#threads};

  \node[itemSty  ] (EW) at (0   ,- 8*\d) {Call AXPY Function};
  \node[itemSty  ] (TB) at (0   ,- 9*\d) {Timing Begin};
  \node[itemSty  ] (ET) at (1*\d,-10*\d) {Call AXPY Function};
  \node[itemSty  ] (TE) at (0   ,-11*\d) {Timing End};
  \node[itemSty  ] (PT) at (0   ,-13*\d) {Print Elapsed Time};

  \draw[edgeSty] (I)  -- (FB);

  \draw[edgeSty] (FB) -- (0*\d,-1.4*\d) -- (1*\d,-1.4*\d) -- (FK);
  \draw[edgeSty] (FK) -- (1*\d,-2.4*\d) -- (0*\d,-2.4*\d) -- (FE);
  \draw[edgeSty] (FE) -- (SV);

  \draw[edgeSty] (SV) -- (ST);
  \draw[edgeSty] (ST) -- (EW);

  \draw[edgeSty] (EW) -- (TB);
  \draw[edgeSty] (TB) -- (0*\d,- 9.4*\d) -- (1*\d,- 9.4*\d) -- (ET);
  \draw[edgeSty] (ET) -- (1*\d,-10.4*\d) -- (0*\d,-10.4*\d) -- (TE);
  \draw[edgeSty] (TE) -- (PT);
\end{tikzpicture} }
  \end{minipage}
  \hfill
  \begin{minipage}[t]{0.48\columnwidth}\vspace{0pt} 
    \hspace{4.8em}\textbf{CUDA}\\[2ex]
    \resizebox{\columnwidth}{!}{ \tikzset{
  itemSty/.style={ellipse, draw, thick, darkgray, fill=backcolor,
    minimum width=4em, minimum height=2ex, align=center, font=\footnotesize},
  itemH1Sty/.style={itemSty, fill=jr@purple!25!white},
  itemH2Sty/.style={itemSty, fill=jr@lightblue},
  opSty/.style={draw, thick, darkgray, fill=jr@lightgreen,
    minimum height=4ex, rounded corners=1ex, align=center},
  edgeSty/.style={->, >=latex, very thick, darkgray}
}
\begin{tikzpicture}
  \def\d{1.0}

  \node[itemSty  ] (I)  at (0   ,- 0*\d) {Include};

  \node[itemH1Sty] (FB) at (0   ,- 1*\d) {Function Begin};
  \node[itemSty  ] (FK) at (1*\d,- 2*\d) {$y_i = a x_i + y_i$};
  \node[itemH1Sty] (FE) at (0   ,- 3*\d) {Function End};

  \node[itemSty  ] (SV) at (0   ,- 4*\d) {Variables};
  \node[itemH1Sty] (ST) at (0   ,- 5*\d) {Set \#threads, \#blocks};
  \node[itemH2Sty] (DA) at (0   ,- 6*\d) {Allocate on device};

  \node[itemH2Sty] (M1) at (0   ,- 7*\d) {Memcopy host to device};
  \node[itemSty  ] (EW) at (0   ,- 8*\d) {Launch AXPY Kernel};
  \node[itemSty  ] (TB) at (0   ,- 9*\d) {Timing Begin};
  \node[itemSty  ] (ET) at (1*\d,-10*\d) {Launch AXPY Kernel};
  \node[itemSty  ] (TE) at (0   ,-11*\d) {Timing End};
  \node[itemH2Sty] (M2) at (0   ,-12*\d) {Memcopy device to host};
  \node[itemSty  ] (PT) at (0   ,-13*\d) {Print Elapsed Time};
  \node[itemH2Sty] (DD) at (0   ,-14*\d) {Deallocate on device};

  \draw[edgeSty] (I)  -- (FB);

  \draw[edgeSty] (FB) -- (0*\d,-1.4*\d) -- (1*\d,-1.4*\d) -- (FK);
  \draw[edgeSty] (FK) -- (1*\d,-2.4*\d) -- (0*\d,-2.4*\d) -- (FE);
  \draw[edgeSty] (FE) -- (SV);

  \draw[edgeSty] (SV) -- (ST);
  \draw[edgeSty] (ST) -- (DA);
  \draw[edgeSty] (DA) -- (M1);
  \draw[edgeSty] (M1) -- (EW);

  \draw[edgeSty] (EW) -- (TB);
  \draw[edgeSty] (TB) -- (0*\d,- 9.4*\d) -- (1*\d,- 9.4*\d) -- (ET);
  \draw[edgeSty] (ET) -- (1*\d,-10.4*\d) -- (0*\d,-10.4*\d) -- (TE);
  \draw[edgeSty] (TE) -- (M2);
  \draw[edgeSty] (M2) -- (PT);
  \draw[edgeSty] (PT) -- (DD);
\end{tikzpicture} }
  \end{minipage}
  \caption{Control flow graphs representing AXPY variants. Left graph represents
    OpenMP variants, and right graph represents CUDA variants.  Nodes with gray
    background are the same across graphs, while purple and blue nodes
    are different.  The purple nodes in the left graph differ between OpenMP-only
    variants (no.\ 1, 3); purple nodes on the right graph differ between
    CUDA-only variants (no.\ 2, 4, and 5).}
  \label{fig:graph-axpy}
\end{figure}

This section demonstrates the use of the entire \cgkit\ tool chain as shown in
\cref{fig:toolchain-cflow-ctree}.
The tool chain starts with \cgkit\ recipes, as proposed in \cref{sec:recipe}, to
describe a sequence of code generation operations.  Executing the recipe will
create a \cgkit\ control flow graph, as introduced in \cref{sec:graph}.  For
brevity we direct show the resulting control flow graphs in
\cref{fig:graph-axpy}, where we limit the presentation to one graph for OpenMP
(left of \cref{fig:graph-axpy}), representing variants 1 and 3 of
\cref{tab:axpy-variants}, and another graph for CUDA (right of
\cref{fig:graph-axpy}), representing variants 2, 4, and 5 of
\cref{tab:axpy-variants}.  We use different colors for the nodes in the graph:
nodes with a gray background are the same in both graphs, and the nodes with
purple or blue backgrounds differ between OpenMP and CUDA.
Furthermore, pertaining to both variants for OpenMP, the purple nodes perform
different code generation operations, namely, inserting code from either of the
the two \cref{alg:axpy-incr-1,alg:axpy-incr-threads}.  Similarly, pertaining to
the three variants for CUDA, the purple nodes differ in the inserted code, which
corresponds to one of the
\cref{alg:axpy-incr-1,alg:axpy-incr-threads,alg:axpy-single-iter}.  For CUDA,
also the thread and block counts of the kernel launch for
\cref{alg:axpy-incr-1,alg:axpy-incr-threads} differ from the number of threads
and blocks used for \cref{alg:axpy-single-iter}; hence, the node ``Set
\#threads, \#blocks'' has a purple background.

From a recipe-generated directed acyclic control flow graph as in
\cref{fig:graph-axpy}, our tool chain next constructs a PST.  The tree levels of
the PST are built via a traversal of the DAG beginning at the graph's root node.
When a node of the graph is visited, the PST is extended with the code
generation operation corresponding to that node.  Compared with the previous
experiment in \cref{sec:axpy-variants-ctree}, where the PST was built manually,
we are now constructing the PST based on the graph.
Doing so allows the PST to be built automatically, and the code
generation operations can become of finer granularity without additional
programming efforts.  This is why a larger degree of code reuse can be reached,
as is demonstrated by the quantity of gray nodes in \cref{fig:graph-axpy}, which
are the same code generation operations across all five variants.

All five variants in this experiment---using the full \cgkit\ tool chain---are
character-by-character identical with the generated code from the experiment in
\cref{sec:axpy-variants-ctree}.  Therefore, the advantages from that section
carry over, too: The generated C/C++ code is human-readable, has consistent
indentation, and is straigthforward to debug; and the compilation works without
errors or warning messages from the compiler.

As before in \cref{sec:axpy-variants-ctree}, we aim to quantify the reduction of
lines of code by counting the lines of C/C++ code in all of the templates and
comparing this number with the lines of generated code.  The resulting absolute
numbers and the relative metric of \emph{C/C++ code reduction} are presented in
\cref{tab:axpy-code-reduc}, bottom row.  We observe that the full \cgkit\ tool
chain results in a code reduction of around 65\,\%, which is an improvement of
about 25\,\% compared with our previous experiment using PSTs only in
\cref{sec:axpy-variants-ctree}.
Note that we intentionally focus on the lines of code of the templates because
for a real application of \cgkit, as in \cref{sec:flashx-variants}, we expect
that the lines of code of recipes or nodes of a control flow graph are
significantly lower than the application's C/C++ or Fortran codes.

\begin{table}[b]
  \caption{Code reduction over all AXPY variants (\cref{tab:axpy-variants}).
    The metric of relative \emph{C/C++ code reduction} is calculated as one
    minus the ratio of input code and generated code.  Note that the generated
    code is character-by-character identical across the two experiments (i.e.,
    two rows).}
  \label{tab:axpy-code-reduc}
  \centering
  \footnotesize
  \begin{tabular}{lccc}
    \toprule
    \thead{\cgkit\ Tools} & \multicolumn{2}{c}{\thead{Lines of Code}} &
    \thead{C/C++ Code} \\[-1ex]
                          & input & generated                         & \thead{Reduction} \\
    \midrule
    PST only                        & 171 & 283 & 39.6\,\% \\
    Recipe, Control Flow Graph, PST & 100 & 283 & 64.7\,\% \\
    \bottomrule
  \end{tabular}
\end{table}

\subsection{Variants for \flashx\ hydrodynamics simulations}
\label{sec:flashx-variants}

\flashx\ has two different units for hydrodynamics solvers, one of them being
Spark~\cite{Couch2021}.  Several variants exist in Spark for dealing with
different characteristics of simulations.
The first kind stems from AMR grid implementation,
where \flashx\ supports two different AMR grid backends,
Paramesh~\cite{macneice2000paramesh} and AMReX~\cite{AMReX_JOSS}.
Both provide block-structured AMR grids for \flashx,
but each has different preferences in updating the solution (i.e., time integration)
and applying a flux correction algorithm that is required by the finite volume
time-stepping scheme.

During the hydrodynamics updates, face-centered fluxes are corrected
at coarse-fine grid boundaries to maintain the solution accuracy for all grid points.
Paramesh assumes that all blocks (i.e., subdomains of the grid with uniform
refined cells) are updated regardless of the levels of refinements of each block;
thus, the flux correction scheme needs to be applied for all levels of refinement at the same time.
On the other hand,
AMReX's primary mode of operation is to
update the AMR blocks level by level,
so the flux correction is required for each level's updates.
Since the flux correction scheme requires data communication among the neighboring blocks,
these two different characteristics of each AMR package demand
two different code structures for the Spark hydrodynamics solver,
even though they have identical numerical algorithms.
High-level versions of the two variants are presented in \cref{lst:spark-all-levels,lst:spark-level-by-level}.

\begin{figure}[tb]
\centering
\begin{lstlisting}[
  language=fortran,
  caption={Solution update and flux correction for all levels at once.},
  label={lst:spark-all-levels}
]
do all_blocks
  ! hydrodynamics updates
end do
call communicate_fluxes() ! p2p communication
do all_blocks
  ! flux correction
end do
\end{lstlisting}

\begin{lstlisting}[
  language=fortran,
  caption={Solution update and flux correction level by level.},
  label={lst:spark-level-by-level}]
do lev = max_level, 1, -1
  call communicate_fluxes() ! p2p communication
  do blocks_on(level = lev)
    ! hydrodynamics updates
    ! flux correction
  end do
end do
\end{lstlisting}
\end{figure}

Another aspect of a Spark solver's variants arises from the Runge--Kutta (RK) time stepper.
Spark adopts the strong stability-preserving Runge--Kutta (SSP-RK) methods~\cite{gottlieb1998total}
for integrating solutions in time with high-order accuracy.
As a multistage method, SSP-RK schemes involve halo exchanges
for the guard cells in every substage updating,
which can be very expensive for a large-scale AMR grid due to irregular
point-to-point communication patterns.
Moreover, halo-exchange costs may dominate the SSP-RK method's computational costs
when a platform presents additional communication or memory movement delays,
(e.g., a heterogeneous system).
To avoid several halo exchanges for advancing a single time step,
we introduced another variant of the SSP-RK method, which we call
the \emph{telescoping mode}.
The idea is to consider additional layers of so-called guard cells for
substage updating.  Therefore,
in each substage, we update the solution \emph{including halo area}
instead of communicating data updated from the neighboring blocks.
As a result, the telescoping version of the SSP-RK method requires
only one communication phase per one full time step but with a thicker halo area.
Although the telescoping mode can reduce the amount of data communication,
it can potentially perform worse than the traditional multistage implementation
depending on the stencil size and the number of data points in a block,
as it requires extra computational costs.
Our recent experiment~\cite{DubeyLeeKlostermanVatai23} on 
RIKEN's Fugaku supercomputer
indicates that the telescoping mode performs better only for very large-scale cases.
We anticipate that the performance gain from the telescoping mode will be further rewarded
in heterogeneous machines since it eliminates host-device data transfers for each substage update.
However, the performance trade-offs from the telescoping mode depend highly
on the simulation size, characteristics, computation intensities, and hardware;
therefore, the best practice would be to conduct the performance analysis ahead of production simulations
and then determine whether to turn on or off the telescoping mode.
To support this practice, the Spark code has to maintain both the telescoping
and non-telescoping implementations simultaneously.
The core differences between traditional (non-telescoping)
and telescoping variants of the RK method are illustrated
in \cref{lst:spark-nontelescoping,lst:spark-telescoping}, respectively, with
simplified codes showing the order computations and communication.

\begin{figure}[tb]
\centering
\begin{lstlisting}[
  language=fortran,
  caption={Traditional (non-telescoping) RK method.},
  label={lst:spark-nontelescoping}
]
do stage = 1, max_stage
  call fill_guardcells() ! p2p communication
  do all_blocks
    ! block initializations
    ! intra stage calculations
  end do
end do
\end{lstlisting}
\begin{lstlisting}[
  language=fortran,
  caption={RK method in telescoping mode.},
  label={lst:spark-telescoping}
]
call fill_guardcells() ! p2p communication
do all_blocks
  ! block initializations
  do stage = 1, max_stage
    ! intra stage calculations
  end do
end do
\end{lstlisting}
\end{figure}

Maintaining each Paramesh/AMReX and telescoping/non-telescoping variant in a separate code involves
significant---but ideally avoidable---pro\-gram\-ming efforts, because all four
variants share the same numerical algorithms
and the differences among them are just the overall code structure.
In our previous study in~\cite{DubeyLeeKlostermanVatai23}
we achieved code unification using macros;
however, the management of macros would become too complex for controlling the
variants at the higher-level granularity as we are attempting in the present work.
We observed that controlling the overall call graphs of static Fortran subroutines using macros
complicates the structures of unified code, involves several duplicated lines,
and causes incoherent code unless inspecting the generated code.
Now, we take the next step in our overall portability solution by utilizing
\cgkit's tool chain with recipes, control flow graphs, and PSTs as described in
\cref{sec:methods}.

While utilizing \cgkit, we keep some adequate use of macros within the
\cgkit\ PST templates and static (nongenerated) code of certain Fortran
subroutines.
One use case for macros,
for example, is to interchange OpenMP directives between
CPU multithreading and GPU target offloading.
Thus, each \cgkit-enabled variant of Spark has another layer of divergent
CPU and GPU versions controlled by the
macroprocessor~\cite{DubeyLeeKlostermanVatai23}, as described in
\cref{tab:spark-all-variants}.
Since these device-specific variants do not differ in the
control flow graph of Spark algorithms,
both the CPU and GPU versions of a Spark variant share the same \cgkit\ recipe.
All possible variants of the Spark solver, using both \cgkit\ and macroprocessor, are presented in \cref{tab:spark-all-variants}.
The variants that we target here have the numbers 1/2, 3, and 5/6, and they are
highlighted in bold font in \cref{tab:spark-all-variants}; our reasoning to
exclude certain variants is given in the table caption.

\begin{table}[b]
  \caption{Overview of all possible variants of Spark solver.
    The variants in the Device column are controlled by the macroprocessor~\cite{DubeyLeeKlostermanVatai23},
    and all other variants are managed by \cgkit\ recipes.
    Variant numbers in bold font are fully supported in \flashx.
    Note that the GPU variant 4 with Paramesh is under development
    and that variants 7 and 8 with AMReX and non-telescoping depend on grid
    infrastructure code that is currently under development.
    (The grid infrastucture belongs to \flashx\ units separate from the Spark
    solver variants we are considering here.)
  }
  \label{tab:spark-all-variants}
  \centering
  \footnotesize
  \begin{tabular}{cccr}
    \toprule
    \thead{Variant} & \thead{Flux correction} & \thead{RK update mode} & \thead{Device} \\
    \midrule
    \textbf{1}& \multirow{4}{*}{All-levels (Paramesh)}   & \multirow{2}{*}{Telescoping}      & CPU \\
    \textbf{2}&                                          &                                   & GPU \\
    \textbf{3}&                                          & \multirow{2}{*}{Non-telescoping}  & CPU \\
            4 &                                          &                                   & GPU \\
    \midrule
    \textbf{5}& \multirow{4}{*}{Level-by-level (AMReX)}  & \multirow{2}{*}{Telescoping}      & CPU \\
    \textbf{6}&                                          &                                   & GPU \\
            7 &                                          & \multirow{2}{*}{Non-telescoping}  & CPU \\
            8 &                                          &                                   & GPU \\
    \bottomrule
  \end{tabular}
\end{table}

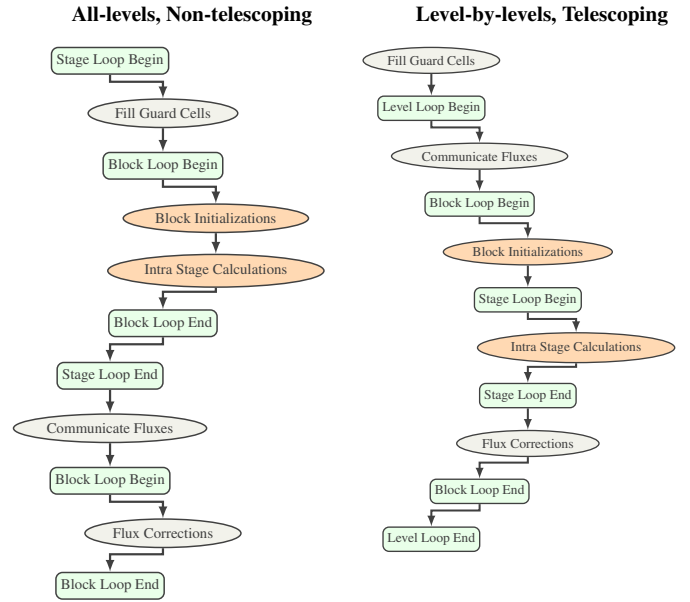
\begin{figure}[tb]
  \footnotesize
  \begin{minipage}[t]{0.48\columnwidth}\vspace{0pt}
    \centering
    \hspace{2em}\textbf{All-levels, Non-telescoping}\\[2ex]
    \resizebox{\columnwidth}{!}{\tikzset{
  commonSty/.style={draw, thick, darkgray, fill=backcolor,
    minimum width=4em, minimum height=4ex, align=center, font=\footnotesize},
  itemSty/.style={commonSty, ellipse},
  itemHlSty/.style={commonSty, ellipse, fill=jr@orange!30!white},
  itemBESty/.style={commonSty, fill=jr@lightgreen, rounded corners=1ex},
  opSty/.style={draw, thick, darkgray, fill=jr@lightgreen,
    minimum height=4ex, rounded corners=1ex, align=center},
  edgeSty/.style={->, >=latex, very thick, darkgray}
}
\begin{tikzpicture}
  \def\d{1.0}

  \node[itemBESty] (SL) at (   0, -0*\d) {Stage Loop Begin};
  \node[itemSty  ] (GC) at (1*\d, -1*\d) {Fill Guard Cells};
  \node[itemBESty] (BL) at (1*\d, -2*\d) {Block Loop Begin};
  \node[itemHlSty] (SI) at (2*\d, -3*\d) {Block Initializations};
  \node[itemHlSty] (IS) at (2*\d, -4*\d) {Intra Stage Calculations};
  \node[itemBESty] (BE) at (1*\d, -5*\d) {Block Loop End};
  \node[itemBESty] (SE) at (0*\d, -6*\d) {Stage Loop End};
  \node[itemSty  ] (CF) at (0*\d, -7*\d) {Communicate Fluxes};
  \node[itemBESty] (BLL) at (0*\d, -8*\d) {Block Loop Begin};
  \node[itemSty  ] (FC) at (1*\d, -9*\d) {Flux Corrections};
  \node[itemBESty] (BEE) at (0*\d, -10*\d) {Block Loop End};

  \draw[edgeSty] (SL) -- (0*\d,-0.4*\d) -- (1*\d,-0.4*\d) -- (GC);
  \draw[edgeSty] (GC) -- (BL);
  \draw[edgeSty] (BL) -- (1*\d,-2.4*\d) -- (2*\d,-2.4*\d) -- (SI);
  \draw[edgeSty] (SI) -- (IS);
  \draw[edgeSty] (IS) -- (2*\d,-4.4*\d) -- (1*\d,-4.4*\d) -- (BE);
  \draw[edgeSty] (BE) -- (1*\d,-5.4*\d) -- (0*\d,-5.4*\d) -- (SE);
  \draw[edgeSty] (SE) -- (CF);
  \draw[edgeSty] (CF) -- (BLL);
  \draw[edgeSty] (BLL) -- (0*\d,-8.4*\d) -- (1*\d,-8.4*\d) -- (FC);
  \draw[edgeSty] (FC) -- (1*\d,-9.4*\d) -- (0*\d,-9.4*\d) -- (BEE);
\end{tikzpicture} }
  \end{minipage}
  \hfill
  \begin{minipage}[t]{0.48\columnwidth}\vspace{0pt}
    \centering
    \hspace{2em}\textbf{Level-by-levels, Telescoping}\\[2ex]
    \resizebox{\columnwidth}{!}{\tikzset{
  commonSty/.style={draw, thick, darkgray, fill=backcolor,
    minimum width=4em, minimum height=4ex, align=center, font=\footnotesize},
  itemSty/.style={commonSty, ellipse},
  itemHlSty/.style={commonSty, ellipse, fill=jr@orange!30!white},
  itemBESty/.style={commonSty, fill=jr@lightgreen, rounded corners=1ex},
  opSty/.style={draw, thick, darkgray, fill=jr@lightgreen,
    minimum height=4ex, rounded corners=1ex, align=center},
  edgeSty/.style={->, >=latex, very thick, darkgray}
}
\begin{tikzpicture}
  \def\d{1.0}

  \node[itemSty  ] (GC) at (   0, -0*\d) {Fill Guard Cells};
  \node[itemBESty] (LL) at (   0, -1*\d) {Level Loop Begin};
  \node[itemSty  ] (CF) at (1*\d, -2*\d) {Communicate Fluxes};
  \node[itemBESty] (BL) at (1*\d, -3*\d) {Block Loop Begin};
  \node[itemHlSty] (SI) at (2*\d, -4*\d) {Block Initializations};
  \node[itemBESty] (SL) at (2*\d, -5*\d) {Stage Loop Begin};
  \node[itemHlSty] (IS) at (3*\d, -6*\d) {Intra Stage Calculations};
  \node[itemBESty] (SE) at (2*\d, -7*\d) {Stage Loop End};
  \node[itemSty  ] (FC) at (2*\d, -8*\d) {Flux Corrections};
  \node[itemBESty] (BE) at (1*\d, -9*\d) {Block Loop End};
  \node[itemBESty] (LE) at (0*\d, -10*\d) {Level Loop End};

  \draw[edgeSty] (GC)  -- (LL);
  \draw[edgeSty] (LL) -- (0*\d,-1.4*\d) -- (1*\d,-1.4*\d) -- (CF);
  \draw[edgeSty] (CF)  -- (BL);
  \draw[edgeSty] (BL) -- (1*\d,-3.4*\d) -- (2*\d,-3.4*\d) -- (SI);
  \draw[edgeSty] (SI) -- (SL);
  \draw[edgeSty] (SL) -- (2*\d,-5.4*\d) -- (3*\d,-5.4*\d) -- (IS);
  \draw[edgeSty] (IS) -- (3*\d,-6.4*\d) -- (2*\d,-6.4*\d) -- (SE);
  \draw[edgeSty] (SE) -- (FC);
  \draw[edgeSty] (FC) -- (2*\d,-8.4*\d) -- (1*\d,-8.4*\d) -- (BE);
  \draw[edgeSty] (BE) -- (1*\d,-9.4*\d) -- (0*\d,-9.4*\d) -- (LE);
\end{tikzpicture} }
  \end{minipage}
  \caption{Simplified graph representations for two variants of Spark algorithms.
    The nodes shown as light green rectangles indicate loop begin and end pairs,
    and nodes shown with gray background denote code generation operations.
    Spark's numerical algorithms are represented in orange, which are subgraphs
    of control flow consisting of multiple nodes, and these subgraphs are reused
    in both variants.
    The subgraphs are encapsulated as functions,
    which take the recipe and nodes as arguments and return the subgraph's final node handle.
    Brief algorithms for each function are described in \cref{alg:spark-block-init,alg:spark-intra-stage}.
  }
  \label{fig:spark-graphs}
\end{figure}

The simplified control flow graph representations for two important Spark variants are depicted in \cref{fig:spark-graphs}.
In our previous study~\cite{DubeyLeeKlostermanVatai23},
all Spark variants were embedded in a single source code, \texttt{Hydro.F90}, using macros.
In that way the \texttt{Hydro.F90} file contained multiple conditional statements
with several duplicated lines for calling internal subroutines,
which are identical to each variant.
Here, however, we use \cgkit\ to handle the overall control flow
for each Spark variant and use macros to manage different versions of Spark's internal subroutines.
Thus, different Spark variants are realized as \cgkit\ recipes
instead of embedded in a single source code using conditional statements.
Maintaining each variant of \cgkit\ recipes has several benefits.
First, it allows simulation developers (e.g., \flashx\ users) to follow the algorithms straightforwardly.
The physics unit developers (e.g., Spark developers)
abstract out lower-level parts---where the abstraction is up to their
choice---into templates, and they expose higher-level parts of the numerical
algorithms with entries in recipes, which correspond to the nodes of the graphs
in \cref{fig:spark-graphs}.
The recipes are realized with just 20 to 30 lines of Python code
(including comment lines), which are more accessible
to \flashx\ users
to track and understand the algorithmic flow of a given physics unit.

Moreover, recipes reduce code duplications and enable more flexible code
composability, which was not possible with previous tools, such as
the macroprocessor.
In Spark's code generation experiments, all variants share two common control
flow subgraphs (orange nodes in \cref{fig:spark-graphs}), which correspond to
essential numerical calculations for Spark.  The benefit of \cgkit\ is that
the subgraphs can be reused in other variants of Spark without duplicating
Fortran source code, which was unavoidable with a unified source code using
macros and conditional rerouting.
The subgraphs can be encapsulated as Python functions,
which take the recipe and a set of required nodes.
The function adds nodes to the recipe in the desired order and returns the final node's handle.
As shown in \cref{fig:spark-graphs}, we encapsulate Spark's core numerical algorithms in subgraphs,
and the functions in \cref{alg:spark-block-init,alg:spark-intra-stage} are used
to produce them in all Spark variants.
Thus, the call graph of Spark's numerical algorithms remains identical to all variants.

\begin{algorithm}[tb]
  \caption{Block initializations for Spark}
  \label{alg:spark-block-init}
  \begin{algorithmic}[1]
  \Function{spark\_block\_init}{recipe, root, nodes}
    \State n = nodes
    \State shockDet $\gets$ recipe.add(n.shockDet)(root)
    \State initSoln $\gets$ recipe.add(n.initSoln)(root)
    \State end $\gets$ recipe.add(n.null)([shockDet, initSoln])
    \Statex \Comment ``null'' node, blocking multiedge, does nothing for PST.
    \State \Return end
  \EndFunction
\end{algorithmic}

\end{algorithm}

\begin{algorithm}[tb]
  \caption{Intra stage calculations for Spark}
  \label{alg:spark-intra-stage}
  \begin{algorithmic}[1]
  \Function{spark\_intra\_stage}{recipe, root, nodes}
    \State n = nodes
    \State grvAccel $\gets$ recipe.add(n.grvAccel)(root)
    \State calcLims $\gets$ recipe.add(n.calcLims)(grvAccel)
    \State calcFlux $\gets$ recipe.add(n.calcFlux)(calcLims)
    \State fluxBuff $\gets$ recipe.add(n.fluxBuff)(calcFlux)
    \State updSoln $\gets$ recipe.add(n.updSoln)(calcFlux)
    \State calcEos $\gets$ recipe.add(n.calcEos)(updSoln)
    \State end $\gets$ recipe.add(n.null)([fluxBuff, calcEos])
    \Statex \Comment ``null'' node, blocking multiedge, does nothing for PST.
    \State \Return end
  \EndFunction
\end{algorithmic}

\end{algorithm}

The generated codes from \cgkit\ are more compact and easier to understand,
since they do not include redundant code lines in never-reached conditionals
dedicated to other variants and thus aid debugging purposes.
For example, each variant generated with \cgkit\ has about 180 lines of code,
but one unified source code contains over 400 lines.
\cref{tab:spark-code-reduc} quantifies the lines of Fortran code inputted to
\cgkit\ via PST templates and the generated line counts; these counts lead to
the metric of relative \emph{Fortran Code Reduction}.  We compare the line
counts and relative reductions of code in three combinations of generated
variants, hence the three rows in \cref{tab:spark-code-reduc}.  The first row
shows the combination of one telescoping and one non-telescoping variant (i.e.,
variants 1/2, 3), which results in around 62\,\% of code reduction.  The second
row considers the two different telescoping variants 1/2 and 5/6, where the
relative reduction is 49\,\%.  The third row considers all of the variants,
where we obtain a code reduction of around 66\,\%.  These are significant
savings of lines of code that will reduce the maintenance efforts of Spark
variants in \flashx.
We note that we achieve a new degree of flexibility for
possible new variants, because one can reuse existing nodes and subgraphs to
construct a new recipe with minimal effort.

\begin{table}[b]
  \caption{Code reduction for Spark variants (\cref{tab:spark-all-variants})
    using \cgkit\ recipe, control flow graph, and PST.
    Three combinations of variants are listed (one per each row), which show
    different configurations where code generation with \cgkit\ is beneficial.
    The metric of relative \emph{Fortran code reduction} is calculated as one
    minus the ratio of input code and generated code.}
  \label{tab:spark-code-reduc}
  \centering
  \footnotesize
  \begin{tabular}{lccc}
    \toprule
    \thead{Spark}    & \multicolumn{2}{c}{\thead{Lines of Code}} &
    \thead{Fortran Code} \\[-1ex]
    \thead{Variants} & input & generated                         & \thead{Reduction} \\
    \midrule
    1/2, 3      & 138 & 360 & 61.7\,\% \\
    1/2, 5/6    & 175 & 343 & 49.0\,\% \\
    1/2, 3, 5/6 & 178 & 525 & 66.1\,\% \\
    \bottomrule
  \end{tabular}
\end{table}

\section{Conclusion}
\label{sec:conclusion}

We presented \cgkit\ as a new solution approach for
code abstraction and code generation for scientific computing.
The  tools in \cgkit\ can be treated as standalone, or they can be combined into
a code generation tool chain.  The major new tools are parametrized source
trees, control flow graphs, and recipes.
Our proposed tool chain that uses all of the three major tools gives
greater control to developers to achieve user-defined abstractions that yield algorithmic
variants with very succinct \cgkit\ recipes in the Python language.
The recipes control the composition of users' PST templates through
which users of \cgkit, such as \flashx\ developers, are able to leverage their
domain knowledge about their code to achieve their desired granularity of code
transformation.

The \cgkit\ tools can manage source code for any programming language generally;
here, we focused on the C/C++ and Fortran languages.
The tools are created with the (human) users playing a central role because of
their domain knowledge, target platform knowledge, and specific workflow aims.
Therefore, generated code is optimized for readability by human programmers.
This allows for straightforward debugging of parallel scientific applications,
thus providing error-free compilation of generated code and correct execution
of an application.  Additionally, we aim for clear input/output relationships
for generated code that allow users to reason about performance and also its
portability across platforms.

The two presented code generation experiments serve two purposes.  One is to
illustrate the usage of the new tools of \cgkit\ while generating variants of a
broadly known operation in numerical linear algebra, the AXPY.  The other
demonstrates a concrete need for generation of variants in the Spark
hydrodynamics solver of the \flashx\ application.
For the AXPY operation, we generate variants that differ in the algorithm and in
the parallelization (OpenMP or CUDA).  The full \cgkit\ tool chain, from recipes
to parsed code, is able to achieve a C/C++ code reduction of 65\,\%.
For the \flashx\ application, variants implement different flux correction
techniques and different algorithms for multistage time stepping.  The proposed
\cgkit\ workflow reduces the Fortran code by up to 66\,\%.  Such reductions
represent a significant advancement in maintainability of complex scientific
codes.  Further, this frees up resources for scientific advancements.

Beyond generation of variants, we have designed \cgkit\ to be part of a new
class of portability solutions, which are grounded in the large-scale
multiphysics application \flashx\ but can be integrated into any other
application.  In future work, we plan to use \cgkit\ tool chains to
integrate our new orchestration runtime, Milhoja~\cite{ONeilWahibDubeyEtAl22},
within \flashx.  \cgkit\ will control the generation of \emph{Milhoja task
functions}, which are functions that are launched on CPU or GPU devices by the
runtime.

\section*{Data availability statement}

The raw data supporting the conclusions of this article will be made available
by the authors upon reasonable request.
The figures of this study are openly available on figshare at
\url{http://doi.org/10.6084/m9.figshare.24922065}, 
\url{http://doi.org/10.6084/m9.figshare.24922077}, 
\url{http://doi.org/10.6084/m9.figshare.24922086}, 
\url{http://doi.org/10.6084/m9.figshare.24922092}, 
and
\url{http://doi.org/10.6084/m9.figshare.24922095}. 

\section*{Funding sources}

This work was supported by
the Exascale Computing Project (17-SC-20-SC), a collaborative effort of the U.S.
  Department of Energy Office of Science and the National Nuclear Security
  Administration, and by
the Scientific Discovery through Advanced Computing (SciDAC) program via the
  Office of Nuclear Physics and Office of Advanced Scientific Computing Research
  in the Office of Science at the U.S.\ Department of Energy under contracts
  DE-AC02-06CH11357 and 
  DE-SC0023472. 

\bibliographystyle{elsarticle-num}
\bibliography{references.bib}

\clearpage
\begin{center}
  \framebox{\parbox{4in}{\scriptsize%
  Government License (will be removed at publication): The submitted manuscript
  has been created by UChicago Argonne, LLC, Operator of Argonne National
  Laboratory (``Argonne").  Argonne, a U.S. Department of Energy Office of
  Science laboratory, is operated under Contract No. DE-AC02-06CH11357.  The
  U.S. Government retains for itself, and others acting on its behalf, a paid-up
  nonexclusive, irrevocable worldwide license in said article to reproduce,
  prepare derivative works, distribute copies to the public, and perform
  publicly and display publicly, by or on behalf of the Government. The
  Department of Energy will provide public access to these results of federally
  sponsored research in accordance with the DOE Public Access Plan.
  http://energy.gov/downloads/doe-public-access-plan.}}
\end{center}

\end{document}